\documentclass[preprint,review,12pt]{elsarticle}
\usepackage{graphics}
\usepackage{amssymb}

\biboptions{sort&compress}

\journal{Carbon}

\begin{document}

\begin{frontmatter}

\title{Structural analysis of polycrystalline graphene systems by Raman spectroscopy}

\author[1,2]{J. Ribeiro-Soares}
\author[3]{M. E. Oliveros}
\author[3]{C. Garin}
\author[3]{M. V. David}
\author[1]{L. G. P. Martins}
\author[3]{C. A. Almeida}
\author[3]{E. H. Martins-Ferreira}
\author[4]{K. Takai}
\author[5]{T. Enoki}
\author[1]{R. Magalh\~{a}es-Paniago}
\author[1]{A. Malachias}
\author[1]{A. Jorio}
\author[3]{B. S. Archanjo}
\author[3]{C. A. Achete}
\author[1,3]{L. G. Can\c{c}ado\fnref{*}}

\address[1]{Departamento de F\'{\i}sica, Universidade Federal de Minas Gerais, Belo Horizonte, MG 30123-970, Brazil}
\address[2]{Departamento de F\'{\i}sica, Universidade Federal de Lavras, Lavras, MG, 37200-000, Brazil}
\address[3]{Divis\~{a}o de Metrologia de Materiais-DIMAT, Instituto Nacional de Metrologia, Qualidade e Tecnologia-INMETRO, Xer\'{e}m, Duque de Caxias, RJ 25250-020, Brazil}
\address[4]{Department of Chemical Science and Technology, Faculty of Bioscience and Applied Chemistry, Hosei University, 372 Kajino-chou, Koganei,
Tokyo 184-8584, Japan}
\address[5]{Department of Chemistry, Tokyo Institute of Thecnology, 2-12-1, Ookayama, Meguro-ku, Tokyo 152-8551, Japan.}

\fntext[*]{Corresponding author. Tel/Fax: +55(31)34095651. E-mail address: cancado@fisica.ufmg.br (L. G. Can\c{c}ado)}

\begin{abstract}
A theoretical model supported by experimental results explains the dependence of the Raman scattering signal on the evolution of structural parameters along the amorphization trajectory of polycrystalline graphene systems. Four parameters rule the scattering efficiencies, two structural and two related to the scattering dynamics. With the crystallite sizes previously defined from X-ray diffraction and microscopy experiments, the three other parameters (the average grain boundaries width, the phonon coherence length, and the electron coherence length) are extracted from the Raman data with the geometrical model proposed here. The broadly used intensity ratio between the C-C stretching (G band) and the defect-induced (D band) modes can be used to measure crystallite sizes only for samples with sizes larger than the phonon coherence length, which is found equal to 32\,nm. The Raman linewidth of the G band is ideal to characterize the crystallite sizes below the phonon coherence length, down to the average grain boundaries width, which is found to be 2.8\,nm. ``Ready-to-use'' equations to determine the crystallite dimensions based on Raman spectroscopy data are given.
\end{abstract}

\end{frontmatter}

\section{Introduction}
\label{s-intro}

Most of the potential applications of graphene as a two-dimensional system are dependent on large area sample production, which can be achieved by the deposition of chemical vapor~\cite{li09a,chen11AA} or exfoliated graphite~\cite{li08b,li08CC}. In both cases, polycrystalline samples are usually obtained, and the key aspects defining the material properties are the atomically-organized crystallite size and the grain boundaries structure~\cite{huang11a,yu11a,kotakoski11a,lee13a,fei13a}. Although the use of Raman spectroscopy as a quick technique to measure the crystallite size ($L_{\rm a}$) of nanostructured graphitic samples is a procedure that has been introduced 45 years ago~\cite{tuinstra70}, the protocols developed up to date are still empirical and dominated by large uncertainties. However, the basis for developing an unified and accurate model for the Raman-based procedure for addressing these key structural aspects are now in place, mostly due to recent work performed on graphene~\cite{lucchese10,beams11a,basko09b,cancado14,beams14,cervenka09}.\\

In 2010, Raman scattering from defects in graphene was used to define the coherence length ($\ell_{\rm A}$) of electrons/holes excited in the visible range~\cite{lucchese10,beams11a}. The results were found in the range of $\ell_{\rm A}$\,$=$\,2\,-\,4\,nm, roughly independent on the excitation laser energy~\cite{cancado11}, consistent with theoretical expectations~\cite{basko09b}. In 2014, near-field Raman scattering in graphene was used to confirm $\ell_{\rm A}$\,$\approx$\,4\,nm~\cite{su13}, and to define the coherence length for optical phonons ($\ell_{\rm C}$), with an observed value of $\ell_{\rm C}$\,$\approx$\,30\,nm~\cite{cancado14,beams14}. Finally, atomically resolved scanning tunneling microscopy (STM) imaging of grain boundaries elucidated the structural aspects on the merging between two misoriented graphene planes~\cite{cervenka09}. This merge region of lateral extension $\ell_{\rm B}$ is a periodic perturbation on the C-C bonding along the boundary axis, necessary to accommodate the connection between two neighboring hexagonal lattices which are not in the same crystallographic orientation~\cite{cervenka09}. This perturbation is characterized by the presence of localized electronic states near the grain boundary, and recent scanning tunneling spectroscopy (STS) measurements showed that the height of these localized states decay exponentially from the grain boundary with a half-decay length of $\approx$\,1.6\,nm, which defines $\ell_{\rm B}\approx$\,3.2\,nm~\cite{cervenka09}.\\

With this information in hand, it is now possible to show how the carbon-carbon stretching (G band at 1584\,cm$^{-1}$) and the disorder induced (D band near 1350\,cm$^{-1}$) spectral features can be used to describe the average size $L_{\rm a}$ of crystallites and the average width of the grain boundaries $\ell_{\rm B}$ in graphene systems. The experimental results and the model are presented in sections~\ref{s-method} and~\ref{s-model}. In section~\ref{s-comp} we elaborate on the novelties of this model as compared to previous research on this topic, demonstrating why the field matured enough to reach an unified model that accounts for crystallites with $L_{\rm a}$ ranging from a few nanometers up to infinity. Besides the model, this development makes it possible to build ``ready-to-use'' formulae for accurate determination of crystallite sizes in polycrystalline graphene systems, which are given in section~\ref{s-summary}.

\section{Experimental Results}\label{s-method}

\subsection{Sample preparation and structural characterization}
\label{s-method1}

The samples were produced by the well-established heat treatment of diamond-like amorphous carbon (DLC)~\cite{cancado06}, which is known to produce graphite nanocrystallites with lateral dimension ($L_{\rm a}$) defined by the heat treatment temperature (HTT)~\cite{ferrari01}. A representative structural image is presented in Figure~\ref{stm}(a), which shows an STM image of a sample with HTT\,=\,2200$^{\circ}$\,C (see the Supplemental Material for experimental details). This image clearly shows that the sample is polycrystalline. Consecutive zooms at the border between two neighboring crystallites are shown in Figs.~\ref{stm}(b) and (c). From the image in Fig.~\ref{stm}(c), the disordered border of thickness $\ell_{\rm B}\approx3$\,nm is clearly seen, with two well-organized hexagonal lattices at each side, corresponding to the atomic structure of two neighboring crystallites. This atomically-resolved image shows that the neighboring crystallites have different lattice orientations. Figs.~\ref{stm}(d,e) show a scanning transmission electron microscopy-bright field image (STEM-BF), and a transmission electron microscopy-dark field image (TEM-DF), respectively, of the sample with HTT\,=\,2300$^{\circ}$\,C. In panel~\ref{stm}(d), the STEM-BF image reveals the presence of Moir\'e patterns generated by rotation between the hexagonal lattices of adjacent layers. The TEM-DF image [panel~\ref{stm}(e)] clearly shows well defined crystallites, and similar images were used to extract the average $L_{\rm a}$ values of selected samples (see details in the Supplemental Material available).\\

\begin{figure}
\begin{center}
\includegraphics[scale=0.7]{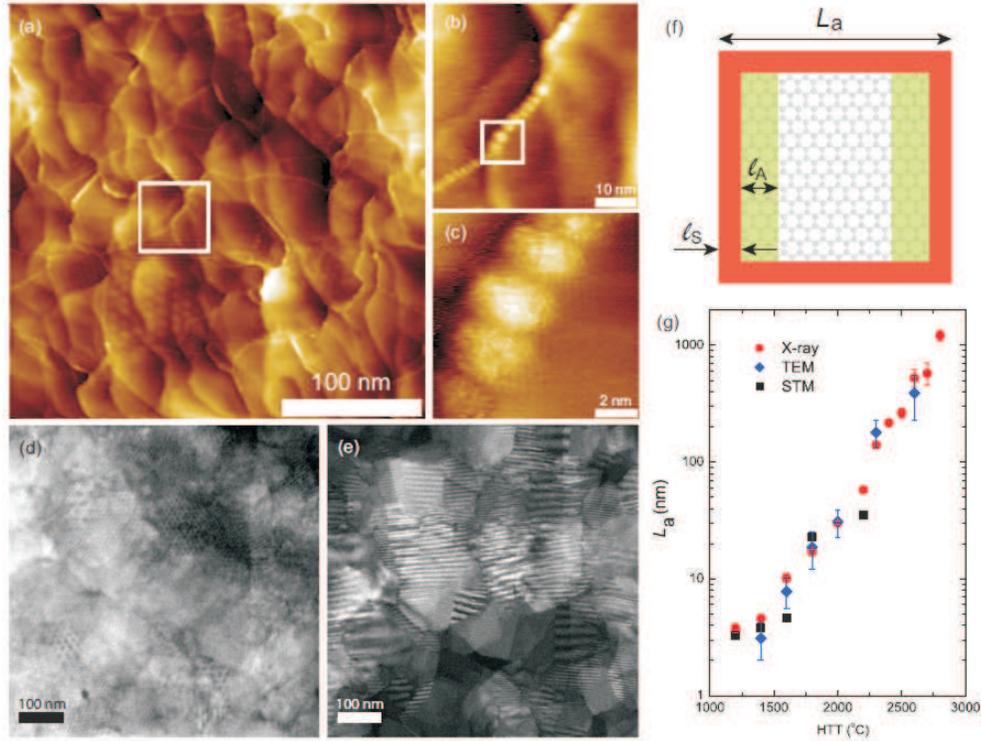}
\caption[] {(a) STM image of the sample with HTT\,=\,2200$^{\circ}$\,C. (b) STM image obtained over the boxed region in panel~(a). (c) Atomically-resolved STM image obtained over the boxed region in panel~(b), showing the disordered border of thickness $\ell_{\rm B}\approx3$\,nm, with two well organized hexagonal lattices at each side, corresponding to the atomic structure of two neighboring crystallites. (d) Scanning transmission electron microscopy-bright field image (STEM-BF), and (e) transmission electron microscopy-dark field image (TEM-DF) of the sample with HTT\,=\,2300$^{\circ}$\,C. Grain boundaries and Moir\'e patterns are observed. (f) Illustration of the idealized crystallite structure: a square-shaped region of side $L_{\rm a}$, formed by a perfect graphene lattice ($\mathbb{A}$ domain) surrounded by the structurally-disordered area (red) of thickness $\ell_{\rm S}$ ($\mathbb{S}$ domain). (g) Summary of the values of crystallite size $L_{\rm a}$ obtained experimentally from different techniques (X-ray diffraction, TEM, and STM) as a function of the heat-treatment temperature (HTT). The experimental details for each technique are provided in the Supplemental Material.\label{stm}}
\end{center}
\end{figure}

The model used to describe the Raman spectral response from the nanocrystallites is illustrated in Fig.~\ref{stm}(f), and will be discussed in section~\ref{s-model}. For this development, twelve different HTT were used to produce polycrystalline graphene with twelve different $L_{\rm a}$ values, which were characterized using X-ray diffraction, transmission electron microscopy (TEM), and STM (see the Supplemental Material for experimental details). These three different techniques were used to accurately measure the mean values of $L_{\rm a}$, and the results are plotted in Fig.~\ref{stm}(g) as a function of the HTT. Most important, the results from the surface technique (STM) are consistent with the results from X-ray and TEM, which probe the volume. This result, together with the fact that the Raman features that will be analyzed here do not change significantly with the number of layers\footnote{The stacking properties of the samples used in this work have been previously investigated by X-ray diffraction and Raman spectroscopy, and the results are reported in Ref.~\cite{cancado08c}. It has been shown that for HTTs below 2300$^{\circ}$\,C ($L_{\rm a}\,<140$\,nm), the samples do not present any detectable stacking order (turbostratic structure). For HTTs\,=\,$2300^{\circ}$\,C and upwards ($L_{\rm a}\,\geq\,140$\,nm) the crystallite size along the $c$ axis ($L_{\rm c}$) increases with HTT. Although the shape of the D band can be slightly influenced by stacking, its relatively low intensity (compared to the G band intensity) obtained for samples with $L_{\rm a}\,\geq\,140$\,nm ensure that the analysis presented here is not significantly affected by the occurrence of stacking order in those particular samples.}, guarantees that our results apply to single and N-layers graphene systems.

\subsection{Raman spectroscopy measurements and analysis}
\label{s-method2}

\begin{figure}
\begin{center}
\includegraphics[scale=0.9]{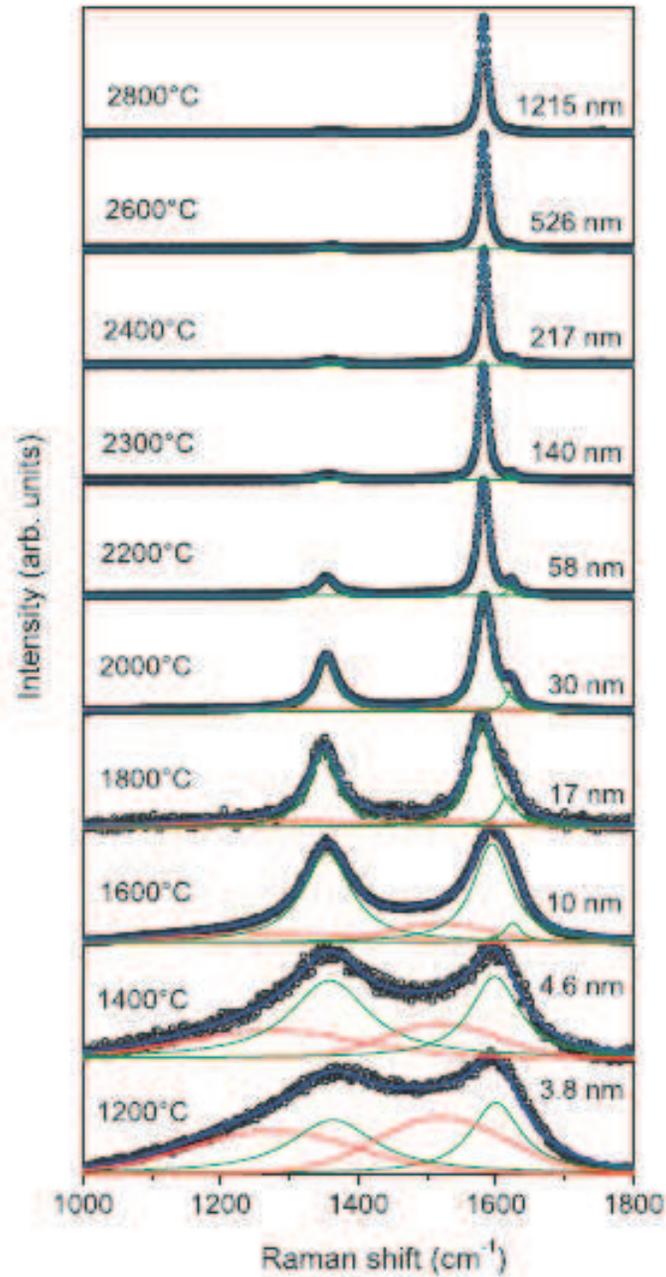}
\caption[] {Representative first-order Raman spectra (empty circles) of heat-treated polycrystalline graphene samples with different crystallite sizes $L_{\rm a}$. The respective values of heat-treatment temperature (HTT) and $L_{\rm a}$ (obtained by X-Ray diffraction) are indicated in the left and right sides of each spectrum, respectively. All spectra were obtained using an excitation laser energy $E_{\rm L}$\,=\,2.33\,eV (wavelength $\lambda_{\rm L}\,=\, 532$\,nm). A detailed description about the procedures to fit the Raman spectra is provided in the Supplemental Material available.\label{espectros}}
\end{center}
\end{figure}

Figure~\ref{espectros} shows representative first-order Raman spectra of heat-treated polycrystalline graphene samples with different different HTT values and correspondingly different crystallite sizes $L_{\rm a}$. The values of HTT and $L_{\rm a}$ (obtained by X-ray diffraction) are indicated in the plot, at the left and right sides of each spectrum, respectively. For samples with HTT\,=\,2200$^{\circ}$\,C and upwards (the five top spectra in Fig.~\ref{espectros}), the Raman spectra are fitted using Lorentzians (green lines). Two main peaks are related to the D and G bands (named here D$^{\mathbb{A}}$ and G$^{\mathbb{A}}$), plus a weak disorder-induced D$^{\,\prime}$ feature at $\sim1610$\,cm$^{-1}$ for the lowest HTT values; for these HTT values the D$^{\,\prime}$ feature is well defined and can be clearly distinguished from the G peak. At HTT\,=\,2800$^{\circ}$\,C, the disorder-induced D and D$^{\,\prime}$ bands are no longer observed. The mechanisms giving rise to the G, D, and D$^{\,\prime}$ peaks have been vastly discussed in the literature, and the details can be reviewed in Refs.~\cite{pimenta2007,malard2009,jorio2011raman,ferrari2013raman,beams2015raman,reich2004raman}. \\

For samples with HTT\,$\leq$\,2000$^{\circ}$\,C (five bottom spectra in Fig.~\ref{espectros}), the spectra are fit using four Gaussian peaks (or five if the D$^{\,\prime}$ band is still noticed as a shoulder in the right side of the G band, \emph{e.\,g.} samples with HTT\,=\,1800 and 1600$^{\circ}$\,C). Of the four Gaussians, two are lined in green, related to the D$^{\mathbb{A}}$ and G$^{\mathbb{A}}$ peaks, which are also observed for HTT\,=\,2200$^{\circ}$\,C and upwards. The other two peaks lined in red are new and are here designated as D$^{\mathbb{S}}$ and G$^{\mathbb{S}}$ peaks. A detailed description about the procedures to fit the Raman spectra is provided in the Supplemental Information.

\section{Theoretical model}
\label{s-model}

A single crystallite is idealized as illustrated in Fig.~\ref{stm}(f): a square-shaped graphene of side $L_{\rm a}$, formed by a perfect graphene lattice ($\mathbb{A}$ domain) surrounded by the structurally-disordered area (red) of thickness $\ell_{\rm S}$ ($\mathbb{S}$ domain). Since two neighboring crystallites share one border of thickness $\ell_{\rm B}$, $\ell_{\rm S}=\ell_{\rm B}/2$. $L_{\rm a}$ and $\ell_{\rm S}$ are the two relevant structural parameters on polycrystalline graphene. There are also two other relevant parameters related to the scattering dynamics, which are represented by the electron and phonon coherence lengths. These dynamic parameters have already been measured using Raman spectroscopy in graphene: $\ell_{A} \approx 3$\,nm for electrons~\cite{lucchese10,beams11a}, and $\ell_{\rm C} \approx 30$\,nm for phonons~\cite{beams14}. These definitions are summarized in Table~\ref{SymbolDef}.\\

\begin{table*}
\begin{center}
{\caption{\label{SymbolDef} Definition of the parameters introduced in the theoretical model.}
\resizebox{\textwidth}{!}{\begin{tabular}{cc}
\hline
\hline
Notation & Definition  \\
\hline
$L_{\rm a}$ & average in-plane crystallite size \\
$\ell_{\rm A}$ & coherence length of electrons/holes \\
$\ell_{\rm C}$ & coherence length of optical phonons \\
$\ell_{\rm B}$ & average width of grain boundaries \\
$\ell_{\rm S}$ & average width of the structurally-disordered area ($\ell_{\rm S}=\ell_{\rm B}/2$) \\
\hline
\hline
\end{tabular}}}
\end{center}
\end{table*}

The importance of $\ell_{A}$ is that this dynamics parameter defines how far from the edge the defect-induced scattering can occur. Therefore, $\ell_{A}$ defines the thickness of the D band scattering within the $\mathbb{A}$ domain, which is the green area in Fig.~\ref{stm}(f). The reason why these areas are considered here in two faces is related to the D band dependence on the laser polarization. The D band scattering is maximum if the polarization of the exciting field is parallel to the edge, and minimum (null for perfect edges) if the exciting field is polarized along the direction perpendicular to the edge~\cite{cancado04,casiraghi09b}. Considering the exciting field as parallel to a pair of opposite edges in the squared crystallite, the D band scattering originates from the two parallel edges only. If the incident light is unpolarized, the D band response would come from the four edges in the square, but the pertinent electric field would be half of the total field, and the same result would be obtained. Considering the pertinent scattering components, the result is actually the same for any crystallite shape.\\

The importance of $\ell_{\rm C}$ is related to spatial confinement, which generates uncertainty in the phonon momentum associated with the finite size of crystallites. If $L_{\rm a} < \ell_{\rm C}$, the Raman-allowed phonon wavevector $q$ is relaxed, leading to the broadening of the Raman bands. Therefore, there is a relation between the crystallite size $L_{\rm a}$ and the width $\Gamma$ of the Raman peaks originated from the $\mathbb{A}$ domain for samples with $L_{\rm a}<\ell_{\rm C}$.\\

The two structural and the two dynamic parameters discussed above can be measured from the $L_{\rm a}$ dependence of the G band width ($\Gamma_{\rm G}$) and from the intensity ratio between the D and G bands ($I_{\rm D}/I_{\rm G}$), as discussed below.

\subsection{The structural ($\ell_{\rm B}$) and dynamic ($\ell_{\rm A}$) parameters versus $L_{\rm a}$: the $I_{\rm D}/I_{\rm G}$ intensity ratio}
\label{s-model2}

\begin{figure}
\begin{center}
\includegraphics[scale=0.6]{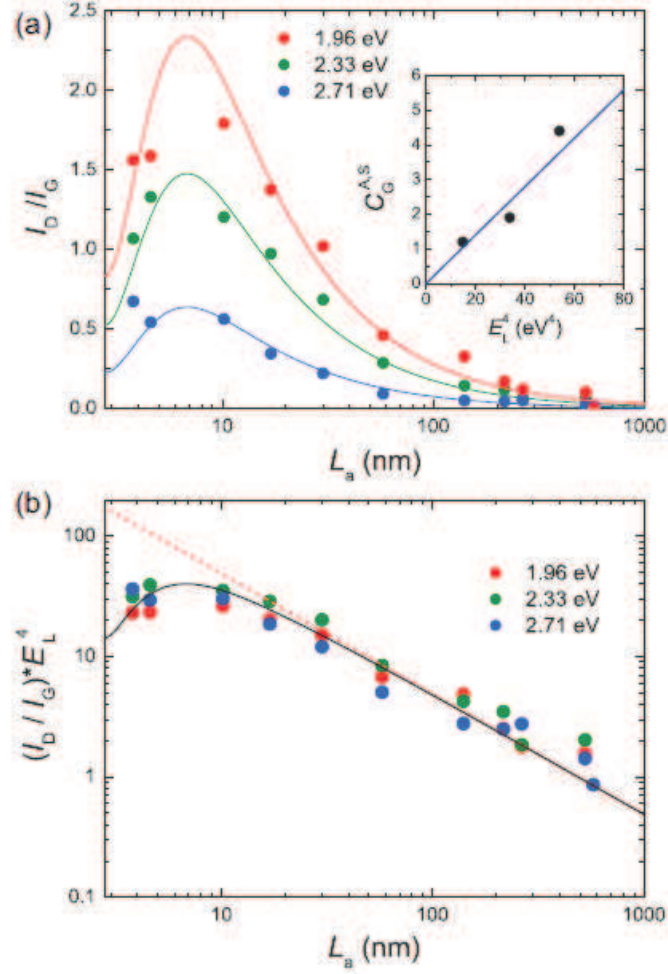}
\caption[] {(a) Plot of the ratio $I_{\rm D}/I_{\rm G}$ as a function of $L_{\rm a}$ for three sets of data obtained using three distinct excitation laser energies, namely 1.96, 2.33, and 2.71\,eV (wavelengths 633, 532, and 458\,nm, respectively). The intensity values correspond to the integration areas. The solid lines are the fitting curves based on Eq.~(\ref{final}). Inset: plot of $C_{\rm G}^{\mathbb{A,S}}$ as a function of $E_{\rm L}^{4}$. The solid line is a linear fit giving $C_{\rm G}^{\mathbb{A,S}}=0.07E_{\rm L}^{4}$ [Eq.~(\ref{CG})]. (b) Plot of the product $(I_{\rm D}/I_{\rm G})*E_{\rm L}^{4}$ as a function of $L_{\rm a}$ for all data shown in panel~(a), in a log plot. The solid line is the corresponding plot obtained from the substitution of Eq.~(\ref{CG}) in Eq.~(\ref{final}). The dotted line is the plot of $(I_{\rm D}/I_{\rm G})E_{\rm L}^{4}=490/L_{\rm a}$, which is here seen to be valid for $L_{\rm a}\geq30$\,nm [see Eq.~(\ref{final03})].\label{ajusteRaman}}
\end{center}
\end{figure}

Figure~\ref{ajusteRaman}(a) shows the plot of the ratio $I_{\rm D}/I_{\rm G}$, obtained from the fitting discussed in section~\ref{s-method2}, as a function of $L_{\rm a}$ for three excitation laser energies, namely 1.96, 2.33, and 2.71\,eV (wavelengths 633, 532, and 458\,nm, respectively). The intensity values plotted in Fig.~\ref{ajusteRaman}(a) corresponds to the peak areas, representing the probability of the whole scattering process. The solid lines shown in Fig.~\ref{ajusteRaman}(a) correspond to the peak fitting curves to the experimental data based on the model discussed below.\\

The measured Raman intensity related to a vibrational mode $\gamma$ can be expressed as sums over a specific two-dimensional scattering domain $\mathbb{C}$ of the form\\
\begin{equation}\label{model01}
I_{\gamma}^{\mathbb{C}}=\frac{\omega_{\rm s}^{4}}{\epsilon_{0}^{2}c^{4}}\,\int_{\mathbb{C}}\,\biggl|\mathbf{G}(\mathbf{r};\omega_{\rm s})\mathbf{\chi}_{\gamma}^{\,\mathbb{C}}(\mathbf{r};\omega,\omega_{\rm s})\,\mathbf{E}(\mathbf{r};\omega)\,\biggl|^{2}\,d^{2}\mathbf{r}\,,
\end{equation}\\
where $\epsilon_{0}$ and $c$ are the free-space permittivity and speed of light, respectively, ${\bf r}=(x,y)$ is the position at the sample plane, $\omega$ and $\omega_{\rm s}$ are the frequencies of the incident and scattered lights, respectively; $\mathbf{G}({\bf r};\omega_{\rm s})$ is the outgoing Green's function which accounts for the whole system, including the scattering and surrounding media, $\mathbf{\chi}_{\gamma}^{\,\mathbb{C}}(\bf{r};\omega_{s},\omega)$ is the Raman susceptibility of a specific vibrational mode $\gamma$ over the domain $\mathbb{C}\in\{\mathbb{A},\mathbb{S}\}$, and ${\bf E}(\bf{r};\omega)$ is the excitation electric field. As an approximation, we have considered the excitation field, as well as the outgoing Green function, to be both uniform over the crystallite area.\\

The Raman scattering responses from both the $\mathbb{S}$ [red area in Fig.~\ref{stm}(f)] and $\mathbb{A}$ (the rest of the square) domains have to be considered to account for the D to G intensity ratio. Therefore, four contributions for the Raman spectrum of polycrystalline graphene have to be used to analyze their relative intensities:
\begin{equation}\label{initial}
\frac{I_{\rm D}}{I_{\rm G}}=\frac{I_{\rm D}^{\mathbb{S}}\,+\,I_{\rm D}^{\mathbb{A}}}{I_{\rm G}^{\mathbb{S}}\,+\,I_{\rm G}^{\mathbb{A}}}\,.
\end{equation}\\
Because both ${\rm D}^{\mathbb{S}}$ and ${\rm G}^{\mathbb{S}}$ bands originate from highly disordered areas, they present lower frequencies (due to the softening of the phonon modes) as compared to the ${\rm D}^{\mathbb{A}}$ and ${\rm G}^{\mathbb{A}}$ bands, respectively. For larger values of $L_{\rm a}$, the $\mathbb{S}$ domain is relatively irrelevant as compared to the $\mathbb{A}$ domain, and the D and G bands are fit with one Lorentzian peak each. For smaller $L_{\rm a}$ values, the $\mathbb{S}$ domain becomes relevant as compared to the $\mathbb{A}$ domain, and the Raman D and G data are fitted with four Gaussian peaks, two peaks for the D$^{\mathbb{S}}$ and D$^{\mathbb{A}}$ bands, and two peaks for the G$^{\mathbb{S}}$ and G$^{\mathbb{A}}$ bands.\\

Within the structurally-disordered domain $\mathbb{S}$, we consider the D and G Raman susceptibilities, $\chi_{\rm D}$ and $\chi_{\rm G}$ respectively, to be independent of the sample position, $\chi_{\rm D,G}^{\mathbb{S}}(\mathbf{r};\omega,\omega_{\rm s})\approx\chi_{\rm D,G}^{\mathbb{S}}(\omega,\omega_{\rm s})$. In this case, the Raman intensities for the D and G bands in the structurally-disordered domain $\mathbb{S}$ of a single crystallite can be readily evaluated from Eq.~(\ref{model01}) to give\\
\begin{equation}\label{DGS}
I_{\gamma}^{\mathbb{S}}=C_{\gamma}^{\mathbb{S}}(4\ell_{\rm S}L_{\rm a}-4\ell_{\rm S}^2)\,,
\end{equation}\\
with $\gamma\in\{{\rm D,G}\}$. The arguments used for deriving this formula are purely geometrical, considering the relative area of the $\mathbb{S}$ domains with
respect to the total area, described as a function of the structural parameters $L_{\rm a}$ and $\ell_{\rm S}$. The constant factors in Eq.~(\ref{model01}) were grouped in Eq.~(\ref{DGS}) in order to have a single constant, namely $C_{\rm D}^{\mathbb{S}}$ and $C_{\rm G}^{\mathbb{S}}$ for the D and G bands, respectively. These constant factors, to which we will refer here as the overall Raman response, account for the $\omega_{\rm s}^{4}$ dependency, the oscillator strength of the electro-phonon and electron-photon interactions, the magnitude of the excitation field, and for the geometry of the collection optics. It is important to notice that Eq.~(\ref{DGS}) is valid only for $L_{\rm a} > 2\ell_{\rm S}$ in polycrystalline graphene, since for $L_{\rm a} < 2\ell_{\rm S}$ the sample is fully disordered and the size dependence makes no more sense, i.e. $I_{\gamma}^{\mathbb{S}}\sim C_{\gamma}^{\mathbb{S}}$.\\

From similar reasoning, the relative intensity for the G band scattering originated from the perfect lattice area $\mathbb{A}$ of a single crystallite is given by\\
\begin{equation}\label{GP}
I_{\rm G}^{\mathbb{A}}=C_{\rm G}^{\mathbb{A}}(L_{\rm a}-2\ell_{\rm S})^{2},
\end{equation}\\
valid in the limit $L_{\rm a} > 2\ell_{\rm S}$, since for $L_{\rm a} < 2\ell_{\rm S}$ there is no $\mathbb{A}$ domain. Finally, the D band intensity over the perfect lattice area is proportional to the green area $\mathbb{A}$ in Fig.~\ref{stm}(f), leading to\\
\begin{eqnarray}\label{DP}
I_{\rm D}^{\mathbb{A}} = C_{\rm D}^{\mathbb{A}}\,\ell_{\rm A}\,(L_{\rm a}-2\ell_{\rm S})\left[1-e^{-2(L_{\rm a}-2\ell_{\rm S})/\ell_{\rm A}}\right]\, .
\end{eqnarray}\\
Again, the $L_{\rm a} > 2\ell_{\rm S}$ limitation applies. Eq.~(\ref{DP}) also comes from geometrical considerations, subject to the fact that the D band scattering in a graphene is strongly localized near the edge [green area in Fig.~\ref{stm}(f)] \cite{casiraghi09b,lucchese10,beams11a,basko09b}. To account for that, the D band susceptibility is modeled to decay exponentially from the border of the $\mathbb{A}$ domain \footnote{This approach is different from that used in Ref.~\cite{lucchese10} for ion-bombarded graphene samples, for which the D band susceptibility was modeled as a step function surrounding the point defect. However, it is physically sound that a smooth exponential decay is more realistic than the abrupt decay described by a step function. Moreover, the use of a step function gives rise to a strong anomaly in the D band intensity for $L_{\rm a}\leq2\ell_{\rm A}$. To overcome this difficulty, stochastic considerations have to be taken into account, which prevents the determination of a simple analytical formula [such as Eq.~(\ref{DP})] in the polycrystalline graphene case.}, with half-decay length $\ell_{\rm A}$ defined by the coherence length of the electron-hole pair involved in the scattering process~\cite{beams11a}.\\

Substitution of Eqs.~(\ref{DGS}-\ref{DP}) in (\ref{initial}) provides the ingredients for the determination of the ratio between the overall intensities of the D and G bands on the form
\begin{equation}\label{final}
\frac{I_{\rm D}}{I_{\rm G}}=\frac{C_{\rm D}^{\mathbb{S}}\,4\ell_{\rm S}(L_{\rm a}-\ell_{\rm S})\,+\,C_{\rm D}^{\mathbb{A}}\,\ell_{\rm A}\,(L_{\rm a}-2\ell_{\rm S})\left[1-e^{-2(L_{\rm a}-2\ell_{\rm S})/\ell_{\rm A}}\right]}{C_{\rm G}^{\mathbb{S}}\,4\ell_{\rm S}(L_{\rm a}-\ell_{\rm S})+C_{\rm G}^{\mathbb{A}}(L_{\rm a}-2\ell_{\rm S})^{2}}\,.
\end{equation}\\

Eq.~(\ref{final}) can be used to fit $I_{\rm D}/I_{\rm G}$ for all different values of $L_{\rm a}$ limited to $L_{\rm a} > 2\ell_{\rm S}$. The best fit to the data [solid lines in Fig.~\ref{ajusteRaman}(a)] was obtained for $\ell_{\rm S}=1.4\pm0.3$\,nm and $\ell_{\rm A}=4\pm0.8$\,nm. Therefore, the structural determination of the average grain boundaries width obtained here ($\ell_{\rm B} = 2\ell_{\rm S}\approx2.8$\,nm) is in very good agreement with the width of the crystallite borders obtained by STM, as shown in Fig.~\ref{stm} and discussed in Ref.~\cite{cervenka09}. Furthermore, the electron coherence length $\ell_{\rm A}\approx4$\,nm is also in excellent agreement with previous experiments performed in ion-bombarded graphene and graphene edges~\cite{lucchese10,beams11a}.\\

It is known that the G band Raman intensity is proportional to the fourth power of the excitation laser energy $E_{\rm L}$, while the D band Raman cross section does not depend on $E_{\rm L}$ \cite{cancado07,klar13}. The inset to Fig.~\ref{ajusteRaman}(a) shows the plot of the fitted values of $C_{\rm G}^{\mathbb{A,S}}$ as a function of $E_{\rm L}^{4}$ ($C_{\rm G}^{\mathbb{A}} = C_{\rm G}^{\mathbb{S}}$). As expected, the behavior is roughly linear, and the data fitting gives\\
\begin{equation}\label{CG}
C_{\rm G}^{\mathbb{A,S}}=(0.07\pm0.02)E_{\rm L}^{4}\,,
\end{equation}\\
where the coefficient is given in units of eV$^{-4}$. The remaining parameters that comes from the data fitting are $C_{\rm D}^{\mathbb{A}} = 7.2\pm2$ and $C_{\rm D}^{\mathbb{S}} = 1.0$. Actually, since all the values we measure here are relative values, we normalized all values to $C_{\rm D}^{\mathbb{S}} = 1.0$. The D band in the activated region is, therefore, 7.2 times stronger than in the structurally disordered domain. Fig.~\ref{ajusteRaman}(b) shows the plot of the product $(I_{\rm D}/I_{\rm G})*E_{\rm L}^{4}$ as a function of $L_{\rm a}$ for all data shown in panel~\ref{ajusteRaman}(a). It is clear from the graphics that the $I_{\rm D}/I_{\rm G}$ data obtained with different excitation laser energies collapse onto the same curve after the data are scaled by $E_{\rm L}^{4}$. The solid line is the plot obtained from the substitution of Eq.~(\ref{CG}) in Eq.~(\ref{final}).

\subsection{Phonon coherence length ($\ell_{\rm C}$) versus $L_{\rm a}$: the G band width}
\label{s-model1}

\begin{figure}
\begin{center}
\includegraphics[scale=0.7]{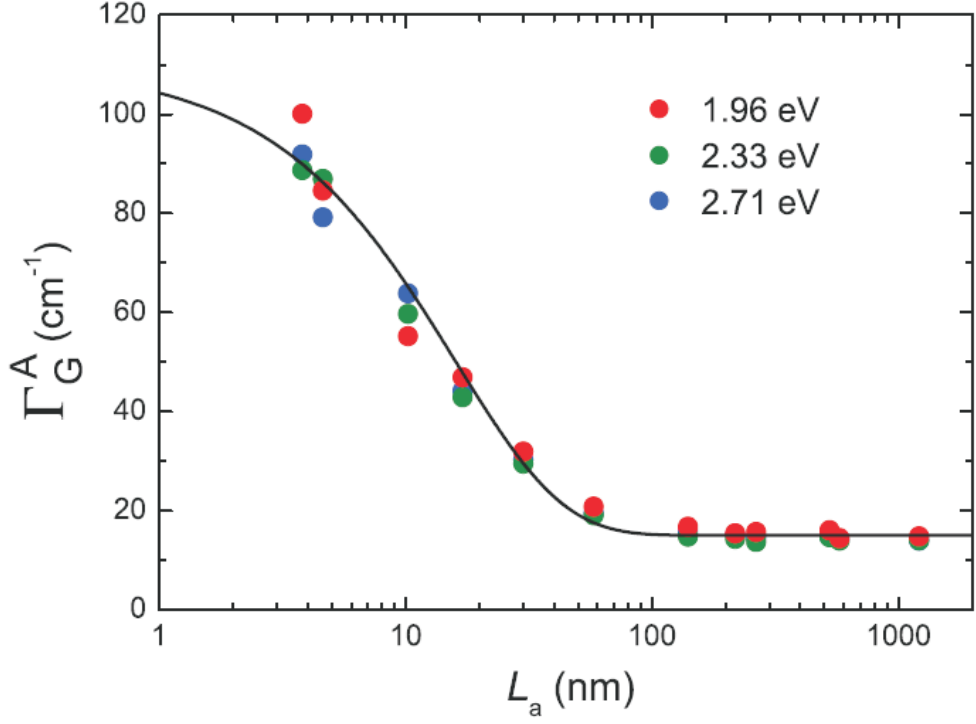}
\caption[] {Plot of $\Gamma_{\rm G}^{\mathbb{A}}$ as a function of $L_{\rm a}$ for the experimental data obtained with three excitation laser sources, namely 1.96, 2.33, and 2.71\,eV (wavelengths 633, 532, and 458\,nm, respectively). The solid line is the fitting according to Eq.~(\ref{eqgamma}).\label{gamma}}
\end{center}
\end{figure}

Figure~\ref{gamma} shows the plot of $\Gamma_{\rm G}^{\mathbb{A}}$ as a function of $L_{\rm a}$ for the experimental data obtained with the three excitation laser sources used here. Referencing to the spectra in Fig.~\ref{espectros}, the peak under consideration is represented by the green line at $\sim 1584$\,cm$^{-1}$. The data do not depend significantly on $E_{\rm L}$, and the whole dataset follows the same trend. For $L_{\rm a}>30$\,nm, we have $\Gamma_{\rm G}^{\mathbb{A}}(\infty)\approx 15$cm$^{-1}$, which is the usual value obtained for undoped pristine graphene~\cite{casiraghi07}. For $L_{\rm a}\leqslant30$\,nm, the spatial confinement of the crystallites affect the phonons, and $\Gamma_{\rm G}^{\mathbb{A}}$ increases exponentially as $L_{\rm a}\rightarrow 0$. In this scenario, the function $\Gamma_{\rm G}^{\mathbb{A}}(L_{\rm a})$ can be approximated as an exponential decay function of the form\\
\begin{equation}\label{eqgamma}
\Gamma_{\rm G}^{{\mathbb{A}}}(L_{\rm a})=\Gamma_{\rm G}^{{\mathbb{A}}}(\infty)\,+\,C\,e^{\,-L_{\rm a}/(\ell_{\rm C}/2)}\,,
\end{equation}\\
where the parameter $C$ is related to the phonon dispersion relation $\omega(q)$, and $\ell_{\rm C}$ gives the full decay length. The solid line in Fig.~\ref{gamma} is the fitting according to Eq.~(\ref{eqgamma}), where the following parameters were obtained: $\Gamma_{\rm G}^{\mathbb{A}}(\infty)=15\pm3$\,cm$^{-1}$, $C=95\pm20$\,cm$^{-1}$, and $\ell_{\rm C}=32\pm7$\,nm. The value obtained here for the phonon coherence length $\ell_{\rm C}$ is, therefore, in excellent agreement with the values obtained from near-field Raman measurements on pristine graphene~\cite{beams14}.\\

Notice that $\Gamma_{\rm G}^{\mathbb{A}}$ is very sensitive to $L_{\rm a}$ for samples with crystallites smaller than $\sim$\,30\,nm. For $L_{\rm a} > \ell_{\rm C}$ the G band is related to the phonon in the center of the Brillouin zone, and it can be fitted with a single Lorentzian with $\Gamma_{\rm G}^{\mathbb{A}}(\infty)=15$\,cm$^{-1}$. For $L_{\rm a} < \ell_{\rm C}$ the G band is a convolution of contributions from different $q$ values, and the Gaussian function is used to fit the data. This explains the change from Lorentzians to Gaussians to fit the data in
Figure~\ref{espectros}. The evolution is monotonic with $L_{\rm a}$, and no major change happens if only Lorentzians or only Gaussians are used to fit all the data (see the details in the Supplemental Information available).

\section{Comparison with related literature}
\label{s-comp}

For those who are knowledgeable about the research in this topic, it is important to compare the results presented here with what has been published in this field over the last 45 years. The ratio between the integrated intensities of the D and G bands ($I_{\rm D}/I_{\rm G}$) have been broadly used to measure the crystallite size $L_{\rm a}$ of nano-structured graphitic samples. The first approach was originally introduced by Tuinstra and Koenig~\cite{tuinstra70}. While the D band intensity scales with the perimeter of the crystallite ($I_{\rm D}\propto L_{\rm a}$), the G band intensity is proportional to the crystallite area ($I_{\rm G}\propto L_{\rm a}^{2}$), so that $I_{\rm D}/I_{\rm G}\propto 1/L_{\rm a}$. Using this simple approach the authors reached the so-called \emph{Tuinstra and Koenig relation} on the form $I_{\rm D}/I_{\rm G} = \kappa / L_{\rm a}$\, , with $\kappa$ being an empirical proportionality constant. Later on, Mernagh and collaborators observed that the the proportionality constant $\kappa$ scales with the excitation laser energy~\cite{mernagh84}, and Ref.~\cite{cancado06} measured $\kappa\approx560E_{\rm L}^{4}$.\\

Ferrari and Robertson noticed that the \emph{Tuinstra and Koenig relation} was no longer valid for samples with a higher degree of disorder~\cite{ferrari01}. The reason presented by the authors was that the totally symmetric vibrational mode giving rise to the D band involves all six atoms in a hexagonal ring (this mode can be seen as a breathing-like mode of the hexagons). On the other hand, the bond stretching mode (with E$_{\rm 2g}$ symmetry) giving rise to the G band only involves a pair of atoms in the graphene unity cell. In a highly-disordered regime, the D mode would be more affected by broken bonds than the G mode, and in this case the proportionality between the $I_{\rm D}/I_{\rm G}$ ratio and $L_{\rm a}$ should be somehow inverted. By performing a back-of-the-envelope calculation, the authors reached the relation $I_{\rm D}/I_{\rm G}\propto L_{\rm a}^{2}$. Ferrari and Robertson then proposed the separation of nano-structured graphitic structures initially in two distinct groups, called stages I and II, happening around $L_{\rm a} \sim 2$\,nm. The authors also proposed a third group (stage III), in which $sp^{\,3}$ sites become very important, and the samples are transformed from amorphous carbon ($a$-C) to tetrahedral amorphous carbon ($ta$-C). However, this is beyond the scope of the present study.\\

A unified approach dealing simultaneously with stages I and II was introduced in Ref.~\cite{lucchese10} for graphene samples with point defects generated by Ar$^{+}$ bombardment. Although the model proposed for graphene samples with point defects in Ref.~\cite{lucchese10} successfully describes the evolution of the ratio $I_{\rm D}/I_{\rm G}$ as a function of the average point defect distance $L_{\rm D}$ for stages I and II, a similar model was missing for nanostructured samples with crystallites of size $L_{\rm a}$. Our work provides such a model, and introduces the information that the correlation length of the optical phonons $\ell_{\rm C}$ is the key factor for the separation between stages I and II, occurring for $L_{\rm a}$ wider and shorter than $\ell_{\rm C}\approx30$\,nm, respectively.\\

For $L_{\rm a}$ shorter than $\ell_{\rm C}\approx30$\,nm, the spectra have to be fit with contributions from both the pristine lattice and the structurally disordered regions, and four main peaks have to be introduced to fit the overall data (except for a fifth peak accounting for the small D$^{\,\prime}$ peak, when needed). Although the introduction of the ${\rm D}^{\mathbb{S}}$ and ${\rm G}^{\mathbb{S}}$ peaks lead to extra complexity, it is strictly necessary for an accurate analysis (to guide the fitting procedure, a detailed description of the key factors is provided in the Supplemental Material). The tentative approach of treating the Raman spectrum of polycrystalline graphitic samples with extra peaks (besides the usual D and G modes) has been previously introduced in Refs.~\cite{bacsa1993,ishida1986,casado1995,mcevoy2012}. However, these works attribute these two extra contributions to different phonon modes activated by disorder. This is not fully accurate since these extra modes cannot be observed in other types of defects, such as the edges of pristine graphene samples in which, indeed, the highly-disordered peaks are not supposed to be observed.

\section{Summary}
\label{s-summary}

Theoretical considerations supported by experimental data were presented to explain the dependence of the Raman scattering signal on the evolution of structural parameters along the amorphization trajectory of polycrystalline graphene systems. There are four parameters, two structural and two related to the scattering dynamics ruling the Raman response. With the crystallite sizes $L_{\rm a}$ previously defined from X-ray diffraction, transmission electron microscopy, and scanning tunneling microscopy experiments, the theoretical model proposed here provided a methodology for measuring the other three parameters, which are the average grain boundaries width $\ell_{\rm B}$, the phonon coherence length $\ell_{\rm C}$, and the electron coherence length $\ell_{\rm A}$. The structural determination of the average grain boundaries width obtained here ($\ell_{\rm B} = 2\ell_{\rm S}\approx2.8$\,nm) is in very good agreement with the width of the crystallite borders obtained by STM~\cite{cervenka09}. The electron coherence length $\ell_{\rm A}=4$\,nm is in excellent agreement with previous experiments performed on ion-bombarded graphene and on graphene edges~\cite{lucchese10,beams11a}. The value obtained for the phonon coherence length $\ell_{\rm C}$ is in excellent agreement with the values obtained from near-field Raman measurements of pristine graphene~\cite{beams14}. Our model is therefore proved accurate, bringing together well established physical concepts into the practical context of structural analysis that can provide support for technological processes.\\

It is now useful to develop practical formulas for the measurement of the crystallite size $L_{\rm a}$ using Raman spectroscopy. The relation between $\Gamma_{\rm G}$ and $L_{\rm a}$ [Eq.\,(\ref{gamma})] can be inverted to measure $L_{\rm a}$ from the recorded $\Gamma_{\rm G}$ Raman data:\\
\begin{equation}\label{eqgamma02}
L_{\rm a}=\frac{\ell_{\rm C}}{2}\ln\left[\frac{C}{\Gamma_{\rm G}^{{\mathbb{A}}}(L_{\rm a})-\Gamma_{\rm G}^{{\mathbb{A}}}(\infty)}\right]\,.
\end{equation}\\
$\ell_{\rm C}$, C, and $\Gamma_{\rm G}^{\mathbb{A}}(\infty)$ are given in section\,\ref{s-method1}. This formula is ideal for measuring $L_{\rm a}$ between 2.8\,nm and 32\,nm. For $L_{\rm a} < 2.8$\,nm, $\Gamma_{\rm G}$ will be related to the degree of disorder in the $sp^{\,2}$ carbon bonds, i.e. the sample becomes strongly disordered and this measurement of $L_{\rm a}$ no longer makes sense.\\

Although Eq.~(\ref{final}) provides a full description for the evolution of the $(I_{\rm D}/I_{\rm G})$ ratio along the amorphization trajectory of nano-graphitic systems, a more practical formulae for the measurement of the crystallite size $L_{\rm a}$ using Raman spectroscopy can rather be based on the regime $L_{\rm a} > \ell_{\rm C}\approx$\,30\,nm. In this limit, we have $L_{\rm a}\gg\ell_{\rm S}$, and Eq.~(\ref{final}) can be simplified to\\
\begin{equation}\label{final02}
\frac{I_{\rm D}}{I_{\rm G}}\approx\left(\frac{C_{\rm D}^{\mathbb{S}}\,4\ell_{\rm S}\,+\,C_{\rm D}^{\mathbb{A}}\,\ell_{\rm A}}{C_{\rm G}^{\mathbb{A}}}\right)\frac{1}{L_{\rm {a}}}\,,
\end{equation}\\
which corresponds, as expected, to the \emph{Tuisntra and Koenig} relation~\cite{tuinstra70}. Substitution of Eq.~(\ref{CG}), together with the numerical parameters presented in section\,\ref{s-method2} leads to
\begin{equation}\label{final03}
L_{\rm a}\,\approx\,\frac{(490\pm100)}{E_{\rm L}^{4}}\left(\frac{I_{\rm D}}{I_{\rm G}}\right)^{-1}\, .
\end{equation}\\
Ref.~\cite{cancado06} gives a similar result, and the small difference ($\sim10\%$) is ascribed here to a further definition of all experimental parameters in the present work. Eqs.\,(\ref{eqgamma02}) and (\ref{final03}) can be used to determine the atomically-organized crystallite sizes in polycrystalline graphene related systems.\\

\textbf{Acknowledgements}
The Authors acknowledge the financial support from CNPq, FINEP, FAPEMIG, FAPERJ, and Inmetro. LGC acknowledges the grant PRONAMETRO (52600.056330/2012).

\bibliographystyle{elsarticle-num}

\bibliography{soares_Raman_graphene_carbon}

\end{document}